\documentclass{egpubl}
\usepackage{pg2017}

\newcommand{\Fref}[1]{Fig.~\ref{#1}}

\newcommand{\Eref}[1]{Eq.~(\ref{#1})}
\newcommand{\Sref}[1]{Sec.~\ref{#1}}
\newcommand{\pdf}{\mathrm{pdf}}
\newcommand{\etal}{\textit{et al}.}

\newcommand{\eg}{\textit{e}.\textit{g}.}
\newcommand{\para}[1]{\noindent \textbf{#1}}

\newcommand{\argmin}{\arg\min}

\usepackage{amsfonts}
\usepackage{amsmath}

\SpecialIssuePaper         % uncomment for final version of Journal Paper, special issue
 \electronicVersion % can be used both for the printed and electronic version

\ifpdf \usepackage[pdftex]{graphicx} \pdfcompresslevel=9
\else \usepackage[dvips]{graphicx} \fi

\PrintedOrElectronic

\usepackage{t1enc,dfadobe}

\usepackage{egweblnk}
\usepackage{cite}

\title[Photometric Stabilization for Fast-forward Videos]%
      {Photometric Stabilization for Fast-forward Videos}

\author[ X. Zhang \& J. Lee \& K. Sunkavalli \& and Z. Wang]
{\parbox{\textwidth}{\centering Xuaner Zhang$^{1,2}$, Joon-Young Lee$^{2}$, Kalyan Sunkavalli$^{2}$, and Zhaowen Wang$^{2}$}
\\
{\parbox{\textwidth}{\centering $^1$University of California, Berkeley, USA, $^2$Adobe Research, USA}}
}
\begin{document}

% uncomment for using teaser
% \teaser{
%  \includegraphics[width=\linewidth]{eg_new}
%  \centering
%   \caption{New EG Logo}
% \label{fig:teaser}
%}

\maketitle
%------------------------------------------------------------------------\maketitle

\begin{abstract}
Videos captured by consumer cameras often exhibit temporal variations in color and tone that are caused by camera auto-adjustments like white-balance and exposure. When such videos are sub-sampled to play fast-forward, as in the increasingly popular forms of timelapse and hyperlapse videos, these temporal variations are exacerbated and appear as visually disturbing high frequency flickering. Previous techniques to photometrically stabilize videos typically rely on computing dense correspondences between video frames, and use these correspondences to remove all color changes in the video sequences. However, this approach is limited in fast-forward videos that often have large content changes and also might exhibit changes in scene illumination that should be preserved. In this work, we propose a novel photometric stabilization algorithm for fast-forward videos that is robust to large content-variation across frames. We compute pairwise color and tone transformations between neighboring frames and smooth these pair-wise transformations while taking in account the possibility of scene/content variations. This allows us to eliminate high-frequency fluctuations, while still adapting to real variations in scene characteristics. We evaluate our technique on a new dataset consisting of controlled synthetic and real videos, and demonstrate that our techniques outperforms the state-of-the-art.
\begin{classification} 
% according to http:http://www.acm.org/about/class/1998
\CCScat{Computer Graphics}{I.3.3}{Picture/Image Generation}{Display algorithms}
\CCScat{Image Processing and Computer Vision}{I.4.3}{Enhancement
}{Smoothing}
\end{classification}
\end{abstract}
%------------------------------------------------------------------------
\section{Introduction}
The ubiquity of mobile cameras and video sharing platforms such as Youtube, Instagram, and Snapchat has made video capture and processing extremely popular.

However, while it is easy to capture videos, viewing and sharing long, unprocessed videos is still tedious. A popular way to compress videos into shorter clips is to fast-forward them, and timelapse and hyperlapse are two appealing techniques to accomplish this; the former handles videos captured using static (or slow-moving) cameras over a long period of time (\eg a day-to-night landscape shown in one minute), while the latter is applied to videos captured by moving (often hand-held) cameras that covers large distances (\eg a hike across the Great Wall summarized in one minute). These videos are created by sampling only a subset of the frames (either uniformly or taking video features into account~\cite{joshi2015real, poleg2015egosampling}). 

\begin{figure}[h!]
\begin{center}
       \includegraphics[width=1.0\linewidth]{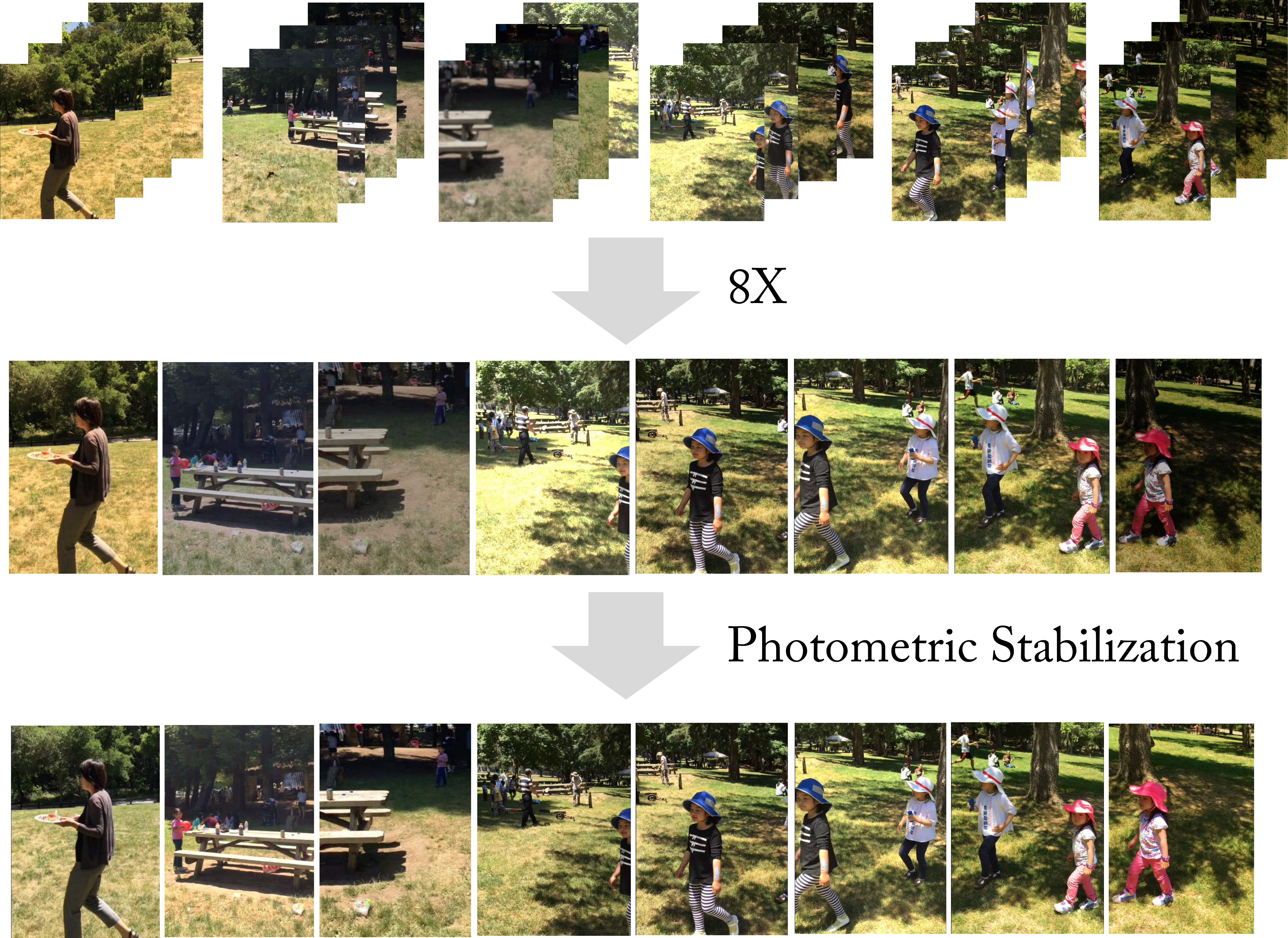}
\end{center}
   \caption{Photometric instability in fast-forward videos. (TOP) original image sequence with photometric jitter, \eg brightness and tone fluctuations. (MIDDLE) sampled frames for fast-forward video with a speedup of 8, which exacerbates the photometric variations (BOTTOM) our result video after photometric stabilization.}
\label{fig:challenge}
\end{figure}

Most videos captured by consumer devices exhibit temporal variations in color and tone that can be caused by either scene changes (\eg variations in scene illumination) and imperfect compensation by in-camera processing such as auto-exposure and white-balance. These photometric fluctuations are particularly troubling when they are high-frequency in nature. This problem is exacerbated in the case of fast-forward videos because frame-sampling changes even low-frequency color and tone variations into bothersome high-frequency flickering (see \Fref{fig:challenge}). Our proposed stabilization framework applies content-aware filtering to fast-forward videos that exhibit undesired photometric jitter and large content variation. We are able to automatically detect and remove high-frequency fluctuations while preserving a smooth scene change.

%------------------------------------------------------------------------

\section{Related Work}
Color and exposure fluctuations are common problems in videos and professional video editing software such as Adobe Premiere and Adobe After Effects have tools to rectify them. However, these tools require significant user effort to manually adjust colors frame by frame. Farbman and Lischinski~\cite{farbman2011tonal} proposed a technique to automatically stabilize tonal variations in videos by applying a pixel adjustment map to align all frames to a set of user-selected anchor frames. Their technique computes dense correspondence by assuming small inter-frame motion, and fails on fast-forward videos which can have large motion and content changes. It also requires the selected frame to have good photometric properties to be a reference. Similar reference-based techniques are also presented in ~\cite{bonneel2013example, bonneel2015blind}, in which they use a video as a filter to transfer the target videos to the same tone and style; in ~\cite{vazquez2014color}, an image is selected as a reference in order to form a consistent set of images taken by various camera sources and settings. In contrast, our method doesn't require a reference video or frame, and is able to handle arbitrary input videos.

Frigo \etal.~\cite{frigo2015motion} extended this work by computing global motion, automatically inserting anchor frames in case of large motion, and weighting the correction by the magnitude of motion. While this improves on the previous technique, it has no notion of content similarity and will fail on videos with large content variation. 
In contrast, our technique computes pair-wise color transformations without requiring dense correspondence, and automatically filters these transformations taking potential content/illumination changes into account.

Wang et al.~\cite{wang2014video} recently propose a stabilization technique that computes pair-wise affine color transformations, which they refer to as color states. They compute "absolute" color transformations between the first frame and subsequent frames; this requires long-range feature tracking that can fail on fast-forward videos. They compute PCA to smooth out the entire sequence states in a synchronized manner. However, their use of PCA over all the color states restricts them to work with only short video clips. In addition, they rely on a frame registration step that only works in the presence of small motion. In contrast, our technique depends on purely local (temporal) processing -- making it computationally more efficient -- and generalizes better to large motion and content change.
%--------------------------------------------------------------------

\section{Photometric Stabilization}
Given a sequence of frames that contain large content variation, our goal is to apply photometric (both luminance and chrominance) stabilization that preserves the original scene illumination changes but removes high frequency color fluctuations.

When the frame sequence has fairly little content change, feature tracking based approaches work quite well~\cite{frigo2015motion, grundmann2011auto}; however, when there are large content variations even within neighboring frames, as often seen in fast-forward videos, feature tracking is not applicable. Thus we rely on only pairwise transformations between successive frames. We accumulate these transformations using regularization to compute longer range transformations. We then smooth these transformations using a temporally-weighted filter that accounts for photometric and content change as well as outlier frames. To avoid artifacts caused by smoothing correlated transformation parameters, we smooth at pixel (correspondence) level, with which to re-compute the desired transformation. Finally, we apply the difference between the original color transformations and their smoothed counterparts to create the final stabilized video.

In the following section, we first describe how we perform photometric alignment between frames then explain how we achieve photometric stabilization over an entire video. 

\subsection{Photometric alignment between frames} 
\label{sec:photo_align}
Given fast-forward videos often have large content variation across neighboring frames, we only calculate the photometric transform between two successive frame pairs. For each pair of frames, we first extract local image features and compute a homography transformation to align the frame pair. We used ORB~\cite{rublee2011orb} feature in all our motion models.

We randomly sample a subset ($5\%$ in our implementation) of corresponding pixel values from the aligned frame pair. We denote the set of sampled correspondences between adjacent frames $i$ and $i+1$ as $(p_{i},q_{i})$. We estimate the pairwise photometric transformation, $T_{i,i+1}$, by minimizing the energy function defined as:
\begin{equation}
\label{eq:pairwiseT}
\sum_{(p_i,q_i) \in P_{i}}{\|T_{i,i+1}(\theta)p_i-q_i\|+\lambda\|T_{i,i+1}(\theta)-\mathbb{I}\|},
\end{equation}
where $(p_i, q_i)$ represents a pair of corresponding pixel values, $\lambda$ is the weight for regularization, and $\mathbb{I}$ denotes an identity transformation.

Our framework does not place constraints on the choice of the transformation model, $T$, to use or color space to work with. In our implementation, we consider photometric smoothing in luminance and chrominance channels separately in the decorrelated \textit{YCbCr} color space. For both luminance and chrominance, we model the color transfer as a global transfer, which is able to account for camera auto-adjustment and global scene illumination change.

When source and target image pair has correspondences, it is more precise to calculate color transfer directly using correspondences, instead of indirectly matching color statistics~\cite{reinhard2001color, pitie2005n, pitie2007automated} or brightness transfer function~\cite{kim2012new}. Although we deal with videos with large content variations, we only compute the transfer between a pair of successive frames, thus the amount of correspondences is sufficient to optimize an accurate pairwise transfer model. Specifically, we use the color transfer model as below.
\vskip 0.1in
\noindent\textbf{Luminance}
We found it sufficient to use a simple weighted gamma curve mapping $T(\{\alpha, \gamma\},Y)$ to model pairwise luminance transfer, which is denoted as: 
\begin{align}
Y^q = \alpha (Y^p)^{\gamma}
\end{align}
where $(Y_q,Y_p)$ are the luminance values at the corresponding pixels, and $\{\alpha, \gamma\}$ are model parameters. 
The effectiveness of gamma curve estimation is described in detail in~\cite{vazquez2015simultaneous}.
\vskip 0.1in
\noindent\textbf{Chrominance}
We use a standard 2$\times$3 affine transformation to model the 2 channel chrominance transfer:
\begin{equation}
\begin{bmatrix}
C_b^q \\
C_r^q \\
\end{bmatrix}
=
\begin{bmatrix}
a_{11} & a_{12} & a_{13} \\
a_{21} & a_{22} & a_{23} \\
\end{bmatrix}
\begin{bmatrix}
C_b^p \\
C_r^p \\
1 \\
\end{bmatrix}.
\end{equation}
We solve for the luminance and chrominance transforms by solving a linear least squares problem (the luminance problem can be made linear by taking log). Since not all correspondences from the alignment are accurate, we add robustness to the step by estimating the parameters via RANSAC. Specifically, for both luminance and chrominance transform, we set maximum iteration of RANSAC to be $1000$, and marked a frame as degenerated when the color transform produces inlier fewer than 1\% of the shorter side of the image size.

Given all the pairwise transformations, $T_{1,2}, T_{2,3},...,T_{N-1,N}$, we can compute the transformation between an arbitrary frame pair $i$ and $j$, by accumulating transformations between them as $T_{i,j} = T_{j-1,j} \cdots T_{i+1,i+2}T_{i,i+1}$.

\begin{figure}[t]
\begin{center}
    \includegraphics[width=0.95\linewidth]{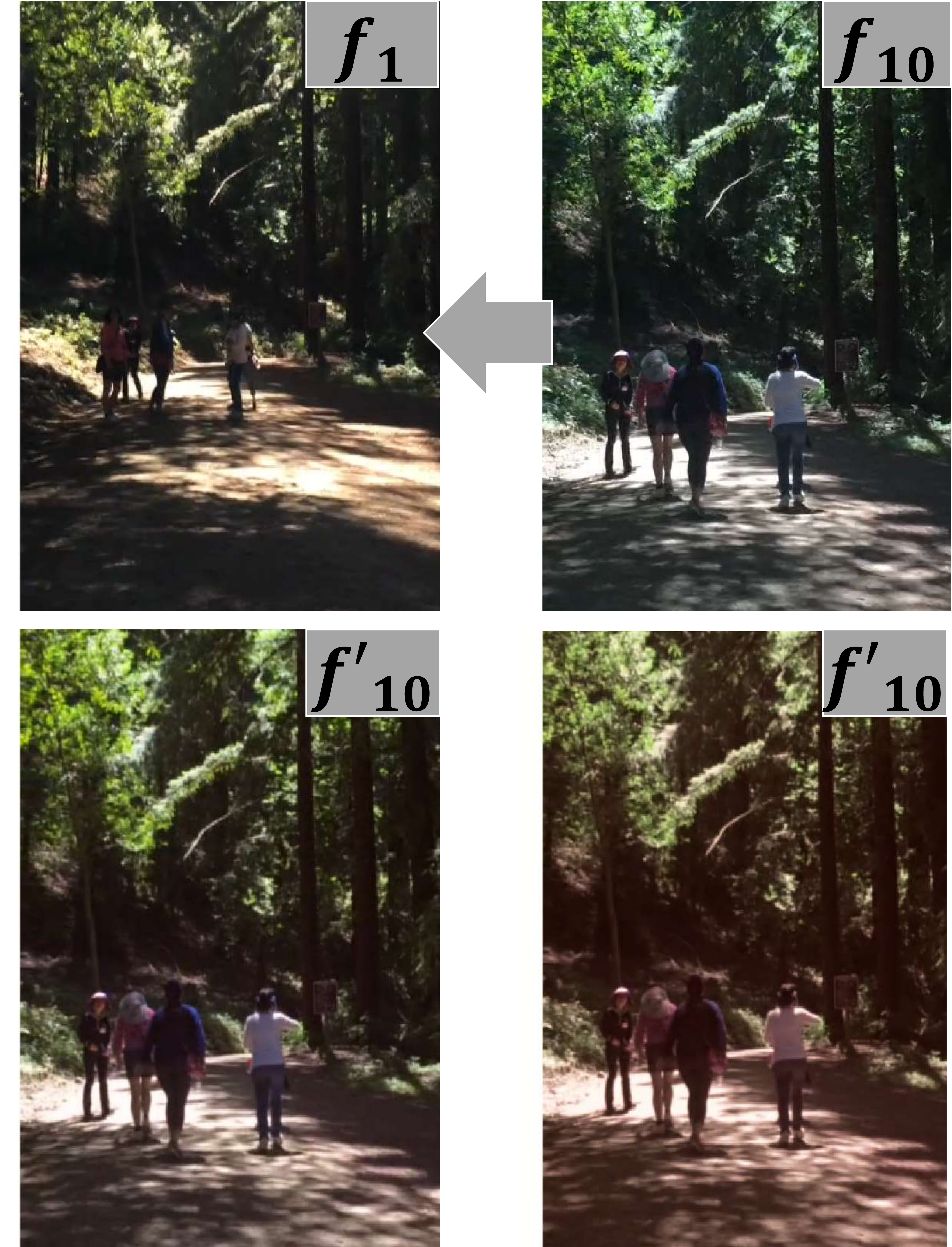}
\end{center}
   \caption{Color transfer by accumulating pairwise transformations. The reference frame is frame $i$ and the target frame is frame $i+10$. The target is matched to the reference using accumulated transformation (in this case, the accumulation of 9 transformation matrices). (TOP) the reference and target frame pair (BOTTOM LEFT) the transformed frame with accumulation (BOTTOM RIGHT) the transformed frame with no accumulation.}
\label{fig:accumulation}
\end{figure}

However, accumulated transformations can introduce color artifacts (see bottom right of \Fref{fig:accumulation} for an example).

To alleviate such model bias, we accumulate correspondences from neighboring frame pairs, (a proportion of $\beta\%$ where $\beta=\frac{100}{2^{|i-k|}}$ such that $k \in [-5,5]$). Thus the pixel samples $(p_{i},q_{i})$ used in \Eref{eq:pairwiseT} are accumulated correspondences from a window of neighboring frame pairs. Note that computing the transformations from features tracked across frames could have been more accurate, but for fast-forward videos, neighboring frames do not have sufficient correspondences.

%--------------------------------------------------------------------

\subsection{Photometric stabilization by weighted filtering}
\begin{figure*}[t]
\begin{center}
   \includegraphics[width=0.95\textwidth]{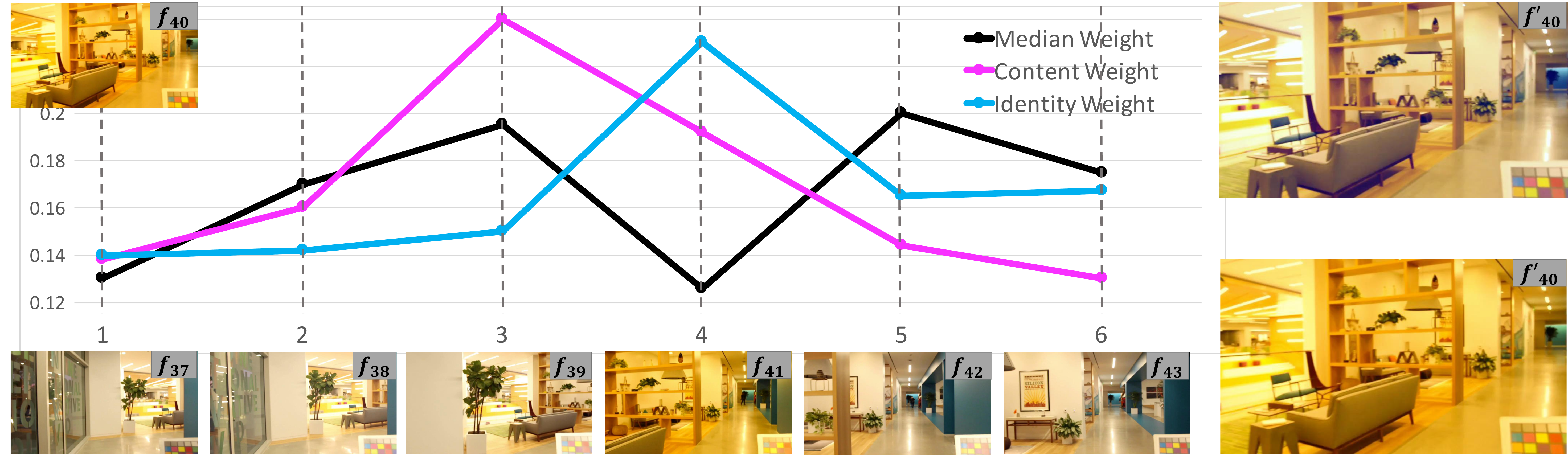}
\end{center}
   \caption{Illustration of the effect of different weight filters. The target frame is plotted on the upper left corner of the figure with the weight plots for all the 6 neighboring frames of the target frame (LEFT). Notice that both the target and one neighboring frame $f_{41}$ are outliers with sharp tone jitter, and $f_{41}$ also contains similar content with the target frame. This leads to high identity and content-aware weights but low outlier weights for $f_{41}$ (because its colors are dissimilar to the remaining frames in this window). Without the outlier weight, $f_{41}$ would be weighted highly among the 6 neighbors leading to poorer filtering (LEFT, BOTTOM), but accounting for it leads to a temporally smoother result (RIGHT, TOP).}
\label{fig:weight}
\end{figure*}

After getting pairwise transformation between arbitrary frame pairs, we filter the transformations to create a set of desired smoothly-varying transformations.

While doing this, it is important to account for content of the video. For example, a large variation in the pixel colors might correspond to a high-frequency jitter in the camera white-balance. On the other hand, it might also be a result of real changes in scene content. Our goal is to remove the first, while smoothly retaining the second. 

To this end, we need a metric that allows us to distinguish between the two. In order to do this, we propose a photometric similarity measure between two frames, that compares their color distributions using the Earth Mover's Distance (EMD)~\cite{rubner2000earth}. Note that earlier when computing pairwise color transfer, we used correspondences-based alignment, which is more robust to outliers (\eg a new foreground object) than histogram transfer. This gives us accurate color transfer but no content information. Here we chose to use histogram-based color comparison between image pairs for similarity measure due to the following two reasons. First, when comparing image pairs with a larger temporal span, correspondence-based approach fails due to the lack of correspondences. Second, histogram-based comparison allows us to infer content variation by comparing color aligned image pairs after correspondences-based alignment. 

Using the EMD measure, we define the photometric distance as:
\begin{equation}
\mathbb{D}_{i,j} = EMD(\pdf(p_i), \pdf(q_j)),
\end{equation}
where $\pdf(p_i)$ and $\pdf(q_j)$ represent the histogram of corresponding pixel values in the frame $i$ and $j$ respectively. Given two frames, we can compute a photometric transformation that aligns the two (as per Sec.~\ref{sec:photo_align}) and then compute the similarity measure. This allows us to eliminate differences that might have been caused by camera adjustments, and then measure content differences.

Given this similarity measure, we now define our weighted smoothing filter. We apply a weighted filter $W_{i}$ of size $M$ to each frame $i$, where $M$ is the number of neighboring frames used to correct the target frame. Denote neighboring frames of frame $i$ as $i-M/2,...,i+M/2$. The overall weight $W$ is composed of four terms: identity weight $W_I$, temporal weight $W_T$, content weight $W_C$ and outlier weight $W_M$. $W$ is computed as a normalized sum of all 4 terms: $W(i,j) = \frac{1}{N_I}W_I(i,j) + \frac{1}{N_T}W_T(i,j) + \frac{1}{N_C}W_C(i,j) + \frac{1}{N_M}W_M(i,j)$, where $N_I$, $N_T$, $N_C$ and $N_M$ are normalization factors computed such that all the weights are at a similar numerical scale. We chose to scale each weight vector by its median. In the following, we will elaborate on the four weights described above.

\para{(1) Identity weight:} The first term is an identity term that penalizes neighboring frames that have different photometric values from the target frame $i$
\begin{align}
W_{I}(i,j)=\exp({-\mathbb{D}_{i,i'}})
\label{eq:identity}
\end{align}
where $i'$ is the simplified notation of transformed pixel samples of the frame $i$ by applying $T_{i,j}$, i.e, $T_{i,j}(p_i)$. If the color of frame $j$ is similar to that of frame $i$, the transform between frame $i$ and $j$ should approach identity. Applying this transformation to $p_i$ should produce values that are very similar to $p_i$. Thus, this metric is a way of evaluating if the neighboring frames are very similar to the current frame (leading to close to identity transformations).

\para{(2) Temporal weight:} The second temporal term simply penalizes frames that are temporally far from the target frame $i$ 
\begin{align}
W_{T}(i,j)=\exp({-((i-j)^2/(2\sigma^2))})
\end{align}

\para{(3) Content-aware weight:} In order to smooth out high-frequency variations, we would like to average out transformations over frames that are similar in content. To do this, we compute the distance between color aligned images, specifically the transformed sample distribution $\pdf(p_i') = \pdf(T_{ij}p_i)$ of frame $i$ and the sample distribution $\pdf(j)$ of frame $j$:
\begin{align}
W_{C}(i,j)=\exp({-\mathbb{D}_{i',j}})
\end{align}
As noted above, differences between the transformed target distribution and the source distribution would indicate content change between the frame pair like dynamic objects in the scene. On the other hand, a simple camera adjustment would get equalized by the transformation $T_{ij}$ leading to larger similarity and a large filter weight. Note that content-aware weight would be equivalent to identity weight when the scene is static. However, when motion presents, content weight penalizes large content change, even if the color transform is close to identity.

\para{(4) Outlier weight:} We define outliers to be frames that contain sharp change in either brightness or color, and should be weighted much less during the computation of reference distribution. However, when computing the filter weights according to measures ($1--3$), such frames will give their neighboring frames low weights (because of the strong changes in color). Instead, we need to eliminate them from the weighting scheme. We assume the outliers are sparse in the selected frame sequence, and thus a majority vote approach such as median filtering would be effective.
\begin{align}
W_{M}(i,j)=\exp({-\|{\mathbb{D}_{i,med}-\mathbb{D}_{i,j}}\|})
\end{align}
where $D_{i,med}$ denotes the EMD distance between the target distribution and the median distribution within its neighboring range. The outlier weight is especially crucial when the target frame is an outlier and there exists other outliers in its neighboring frames.

\begin{figure}[t]
\begin{center}
       \includegraphics[width=1.0\linewidth]{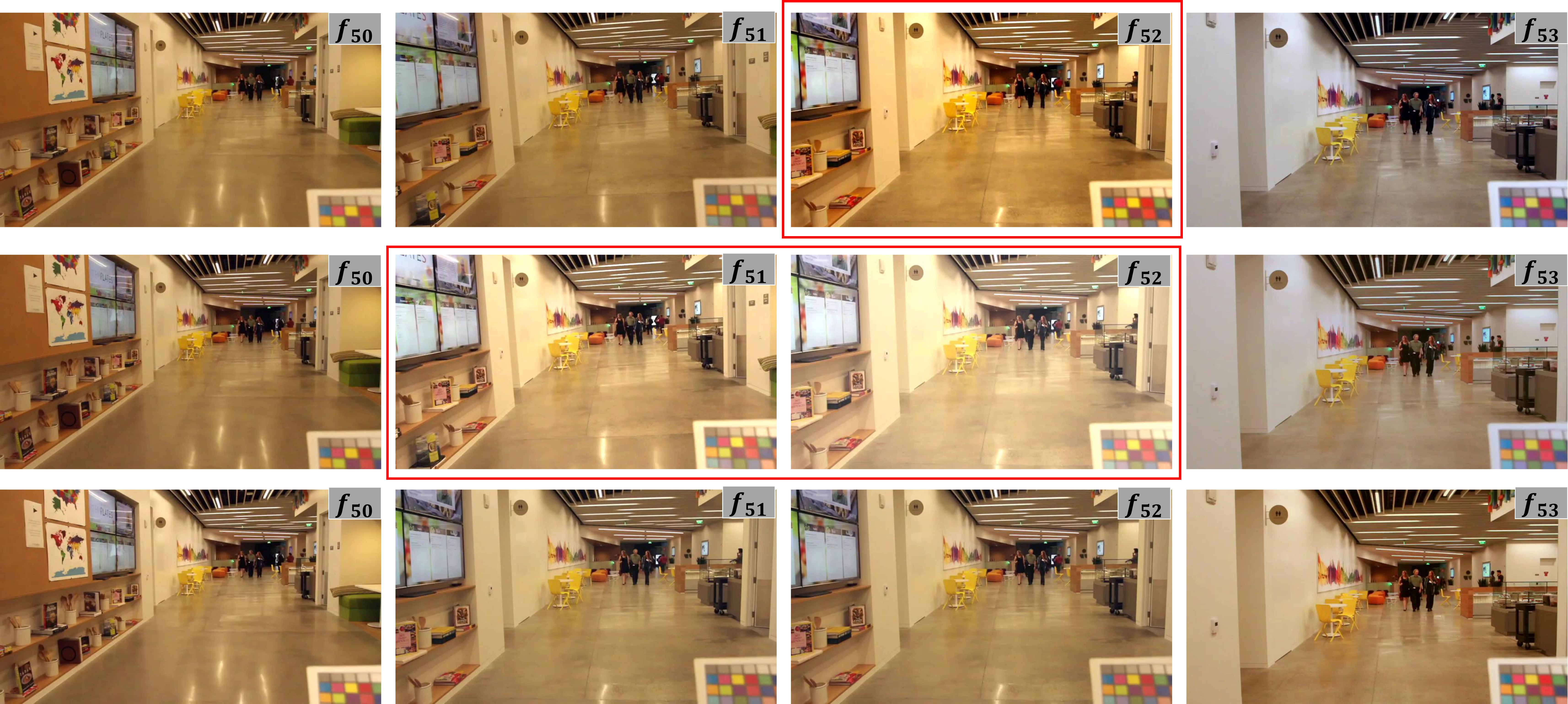}
\end{center}
   \caption{Illustration on the effectiveness of the proposed weighted filtering. (TOP) input sequence with high frequency photometric jitters (\eg the $f_{52}(3^{rd})$ contain brightness jitter, and $f_{53}(4^{th})$ frame contains color fluctuation); (MIDDLE) result of uniform filtering, note that frames with jitter cannot be corrected, and can even affect neighboring frames that are originally correct (\eg the $f_{51}(2^{nd})$ frame is affected by $f_{52}(3^{rd})$ frame to appear brighter) (BOTTOM) result using the proposed content and outlier-aware weighted filtering; the frames can be corrected properly.}
\label{fig:weight_use}
\end{figure}

\Fref{fig:weight} demonstrates how all these weights combine to give us smooth, robust, filtering on a video sequence with jitter. We also compared between using the proposed weighted filtering and using na\"{i}ve uniform filtering; the comparison result is shown in \Fref{fig:weight_use}.

%--------------------------------------------------------

\subsubsection{Rendering photometrically stabilized frames}
Given the transformation weights, we would like to use them to smooth out the photometric variations in the video sequence. One option to do this could be to compute smoothly varying transformations $\hat{T}_{i}$ by directly applying the weighted filter to the original transforms:
\begin{align}
\hat{T}_{i} = \sum_{j=i-M/2}^{j=i+M/2}W(i,j)T_{i,j}.
\end{align} 

However, we notice that directly applying the weighted filters to the transformation matrices results in color artifacts. These color artifacts usually are caused by different components of transformations being filtered independently and thus asynchronously; for example, the $6$ independent variables in affine transformations should not be filtered independently. Wang \etal~\cite{wang2014video} tried to address this problem by smoothing in the PCA-encoded transformation space, but PCA decomposition and their use of only one principal component is limited to short video clips without much content change.

We address this issue by applying the weighted filter $W$ on pixel values instead of on transformation parameters, and then re-compute the desired transformation from the filtered pixel values. Specifically, we take the correspondence points $p_{i}$ from the target frame $i$, transform its distribution to match each of its neighbor frames' distributions, and apply the weighted filter to get the desired color distribution as:
\begin{align}
\hat{p}_{i}=\sum_{j=i-M/2}^{j=i+M/2} W(i,j)\bigg(T_{i,j}p_{i}\bigg).
\label{eq:weighted}
\end{align}

These weighted color values represent the desired smoothly varying pixel values. We then compute a single transformation that aligns the original pixel values, $p_i$ to these weighted pixel values as:

\begin{align}
\argmin_{\theta}\|\hat{T}(\theta)_{i}P_{i}-\hat{P}_{i}\|
\end{align}
where $\hat{P}_{i}$ is the desired distribution calculated in the previous step. $\hat{T}(\theta)$ is then applied to the entire frame to get $i^{'}=\hat{T}(\theta)i$, where $i'$ is the corrected frame $i$. Correcting the video frames via this two-step process leads to results are more robust and artifact-free.
%---------------------------------------------------------------

\section{Photometrically Stable Frame Sampling}
\label{sec:frame_sampling}
\begin{figure}[t]
\begin{center}
   \includegraphics[width=0.95\linewidth]{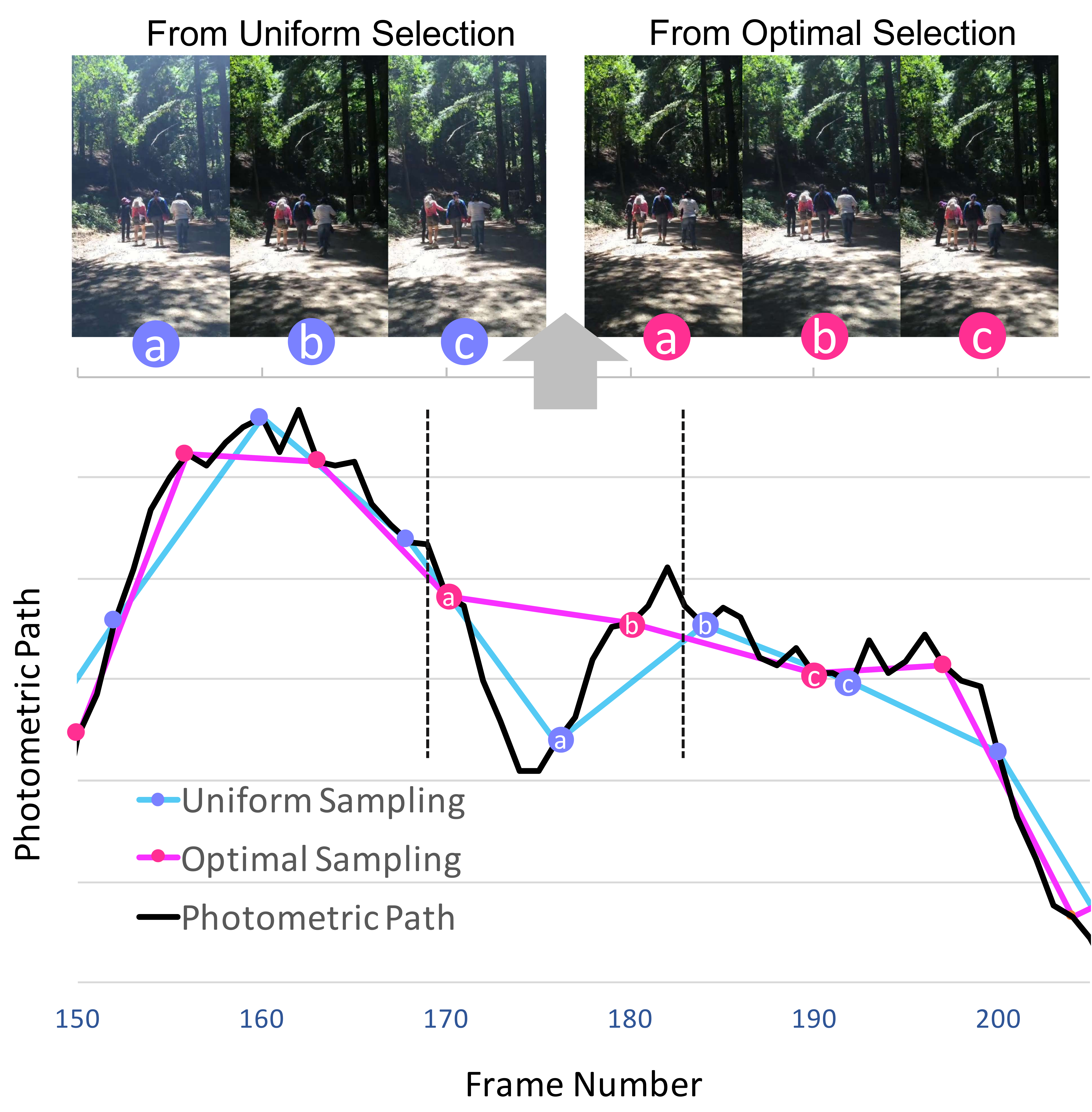}
\end{center}
   \caption{Uniform selection VS. optimal frame selection, which considers photometric constraints. The photometric path is computed using $\bar{P}$ as described in \Eref{eq:pcost}. Note that the path from optimal selection avoids high frequency photometric change and is smoother than either the original or uniform selected paths. Shown on top, We can also visualize the frames being selected using uniform and optimal selection. The optimally selected frames are more photometrically consistent.}
\label{fig:frameselect}
\end{figure}

When generating fast-forward videos, the choice of frames can lead to different stabilized videos. Instead of applying uniform frame selection, we should select frames based on photometric constraints which can help skip degenerate frames such as over-exposed or under-exposed ones, and produce better stabilization. Our frame-sampling technique is similar to Joshi~\cite{joshi2015real}; we define a novel binary photometric and unary blurriness cost and compute the optimal frame sampling by using dynamic programming while maintaining a user-specified frame sampling. The novel photometric costs we introduce are:

\para{Photometric Cost:} During frame sampling we want to remove degenerated frames that have large color changes and information lost (e.g. highly saturated or darkened). We define a simple photometric cost that characterizes the global image temperature and brightness: 
\begin{equation}
  C_{p}(i, j) = \| \bar{P_i} - \bar{P_j}\|
  \label{eq:pcost}
\end{equation}
where the $\bar{P_i}$ and $\bar{P_j}$ denotes the mean correspondence value in frame $i$ and $j$, in the \textit{YCbCr} color space.

\para{Blurriness Cost:} To quantify the blurriness of each frame, we apply a Laplacian kernel to feature patches of each frame and compute the summed variance within the patches. The blurriness cost penalizes frames that contain large motion blur, either from camera motion or dynamic objects in the scene.

These costs can be added to the geometric costs introduced by \cite{joshi2015real} and used to sample optimal frames from a video sequence. \Fref{fig:frameselect} shows how sampling using our photometric costs can lead to more desirable stabilized videos.

%---------------------------------------------------------------------
\section{Experiments}
\begin{figure*}[t]
\begin{center}
	\includegraphics[width=1.0\textwidth]
            {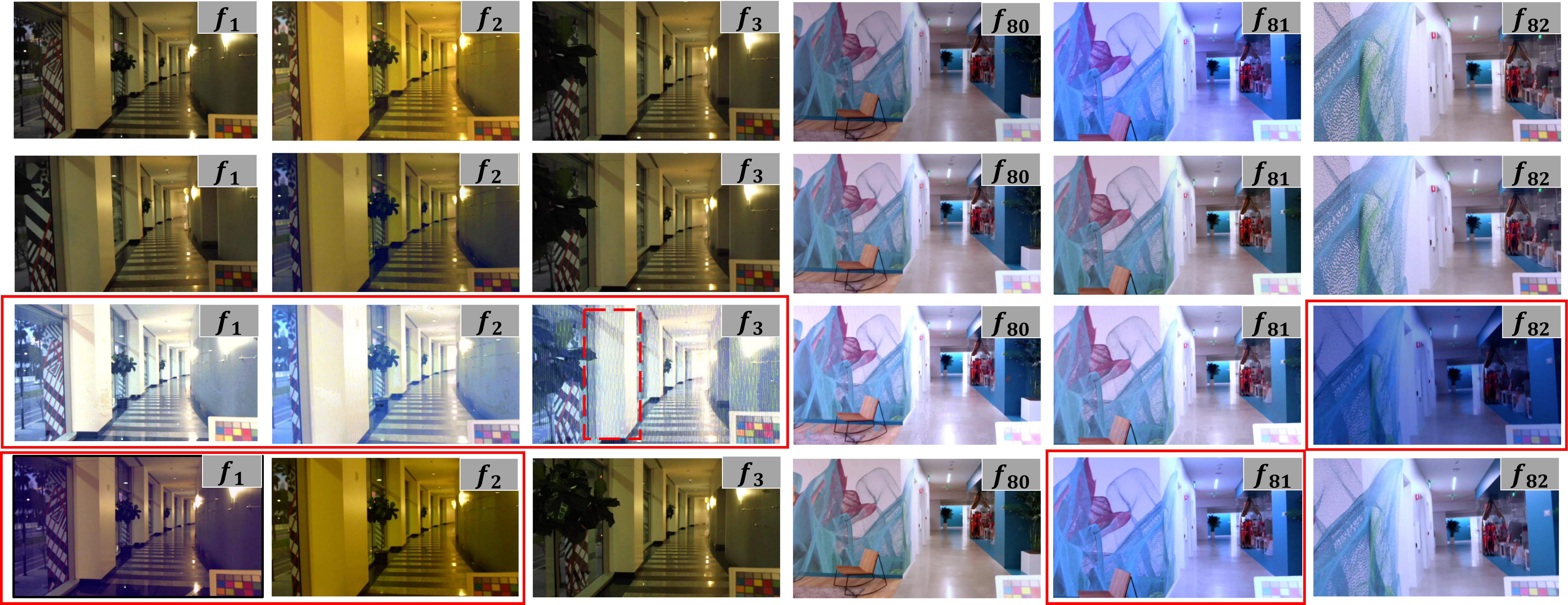}
\end{center}
   \caption{Synthetic experiment on indoor scene dataset. (ROW 1) input sequence with high frequency photometric jitters (\eg the $f_2$ frame contains sharp brightness change, and the $f_{82}$ frame has incorrect white-balance; (ROW 2) stabilized frames using our proposed technique; (ROW 3) stabilized frames using Farbman~\etal~\cite{farbman2011tonal}, in which they try to match the color of all frames to the select anchor frame. Also, local artifacts can be seen in the output due to up-sampling of their adjustment maps; (ROW 4) result of the global affine smoothing method; global smoothing cannot compensate the jitter of $f_{81}$, and can introduce color artifacts as seen in $f_1$.}
\label{fig:synthetic_adobe}
\end{figure*}

\begin{figure*}[t]
\begin{center}
   \includegraphics[width=1.0\textwidth]{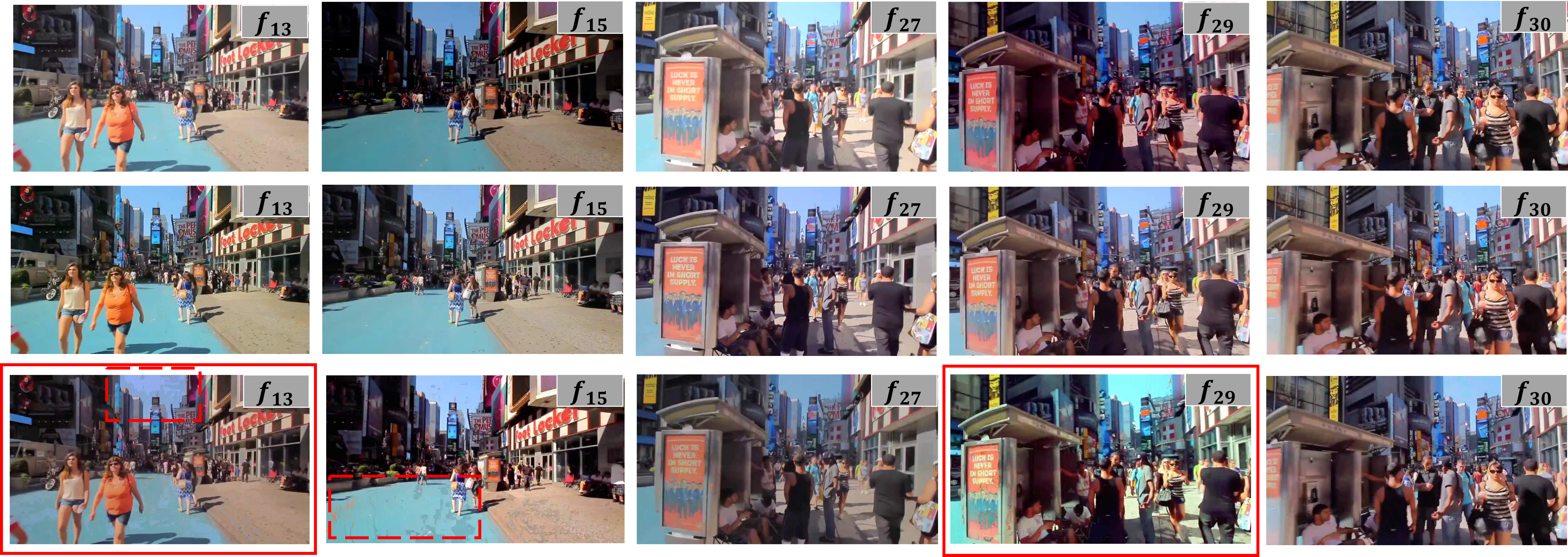}
\end{center}
   \caption{Synthetic experiment on outdoor scene dataset. (ROW 1) input sequence with brightness and color photometric instability (\eg the $2^{nd}$ and $3^{rd}$ contain brightness jitter, and $4^{th}$ frame contains color fluctuation); (ROW 2) stabilized frames using our stabilization method; (ROW 3) stabilized frames using Farbman~\etal~\cite{farbman2011tonal}.}
\label{fig:synthetic_ts}
\end{figure*}

\begin{figure*}[!htbp]
\begin{center}
   \includegraphics[width=1.0\textwidth]{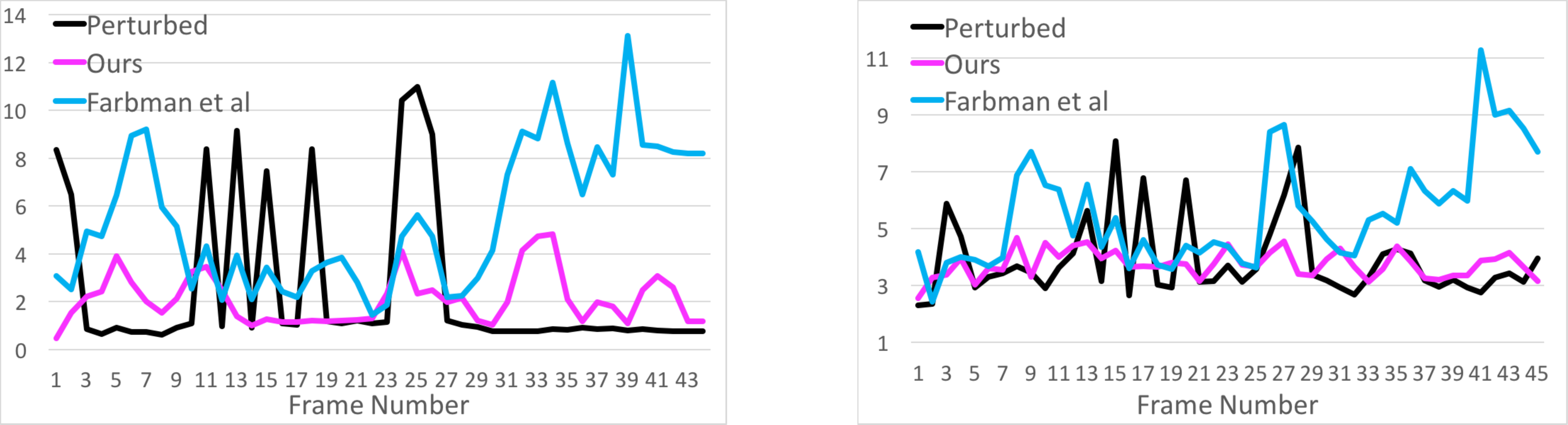}
\end{center}
   \caption{Quantitative evaluations on synthetic video experiments. We compute the RMSE for each frame, between the stabilized video and the ground truth video, and compared with Farbman~\etal~\cite{farbman2011tonal}. (LEFT) result on synthetic indoor scene, corresponds to ~\Fref{fig:synthetic_adobe} (RIGHT) result on synthetic outdoor scene, corresponds to ~\Fref{fig:synthetic_ts}.}
\label{fig:quant}
\end{figure*}

\para{Synthetic videos} To create the synthetic data, we first captured carefully controlled videos with a Canon~7D DSLR camera with locked manual camera settings. Therefore the original frames do not have any photometric variation introduced by camera settings, and brightness and color changes are caused  by scene and illumination changes. We then applied high-frequency tone and color transformations to these videos. Specifically, high frequency jitter is created in the following two ways: 1) apply randomized color transform using our color model described in \Sref{sec:photo_align} 2) manipulate individual frames in existing video editing software to insert temporal inconsistency. These altered video frames were then used to evaluate a variety of stabilization methods (see results on our self-captured video \Fref{fig:synthetic_adobe} and a publicly available video (photometrically stabilized) \Fref{fig:synthetic_ts}). It is possible to quantitatively evaluate on synthetic experiments given ground truth videos that are photometrically stable (for real videos there is no well-defined ground truth and thus we did not perform quantitative evaluation). We compute the RMSE for each frame with the ground truth video, and the results are shown in \Fref{fig:quant}.

\begin{figure}[h!]
\begin{center}
   \includegraphics[width=1.0\linewidth]{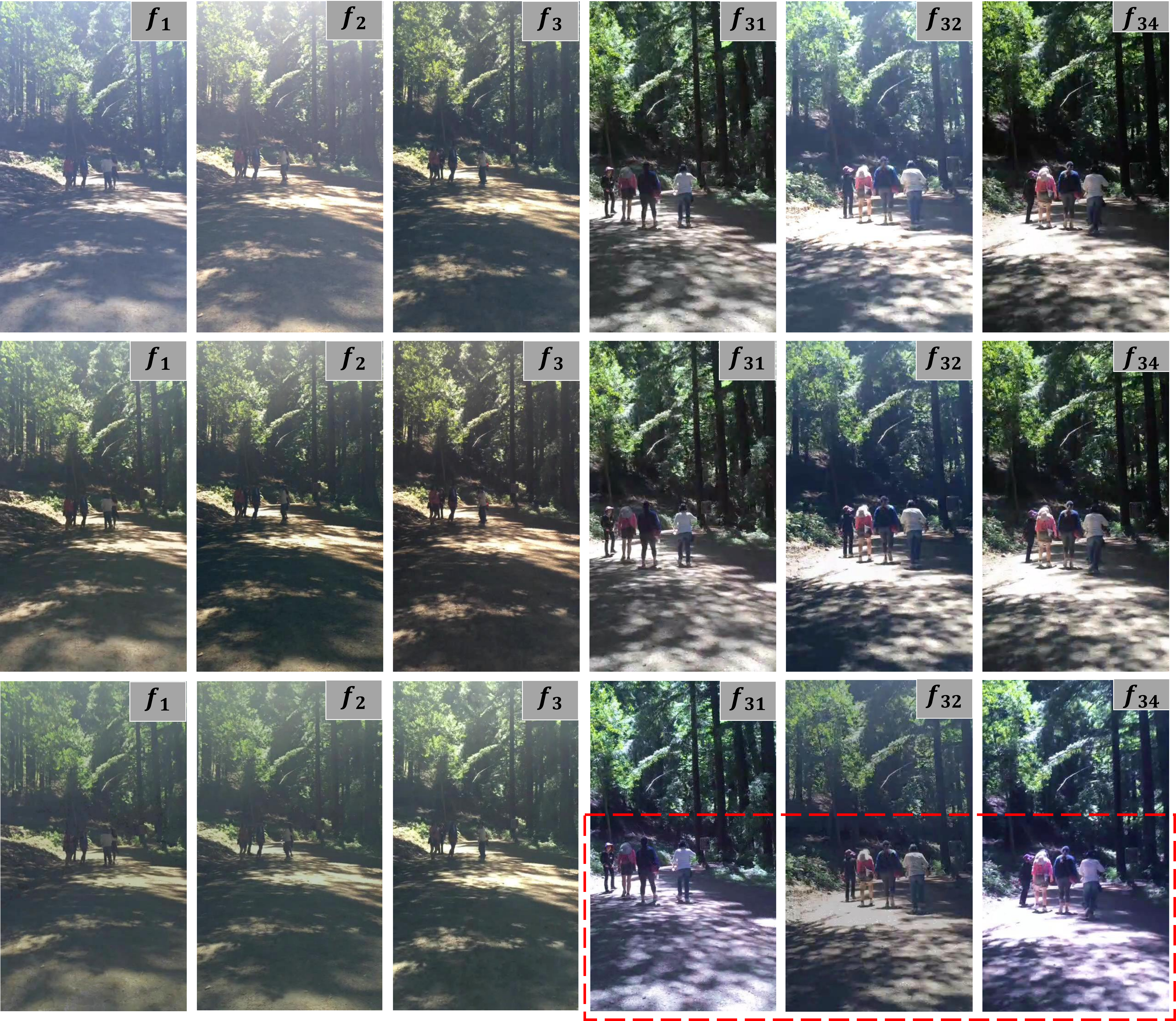}
\end{center}
   \caption{Experiment on a real video dataset. (TOP) input frame sequence; (MIDDLE) corrected frames using our stabilization method; (BOTTOM) corrected frames using Farbman~\etal~\cite{farbman2011tonal}. In the input frame sequence, $f_1$ contains tone jitter, and $f_1, f_2, f_{32}$ contain high frequency brightness fluctuation; the method from Farbman \etal~\cite{farbman2011tonal} does not take into account the scene illumination change, and also introduces artifacts for frames such as $f_{31}$ and $f_{34}$.}
\label{fig:result_hike}
\end{figure}

\begin{figure*}[h!]
\begin{center}
   \includegraphics[width=1.0\textwidth]{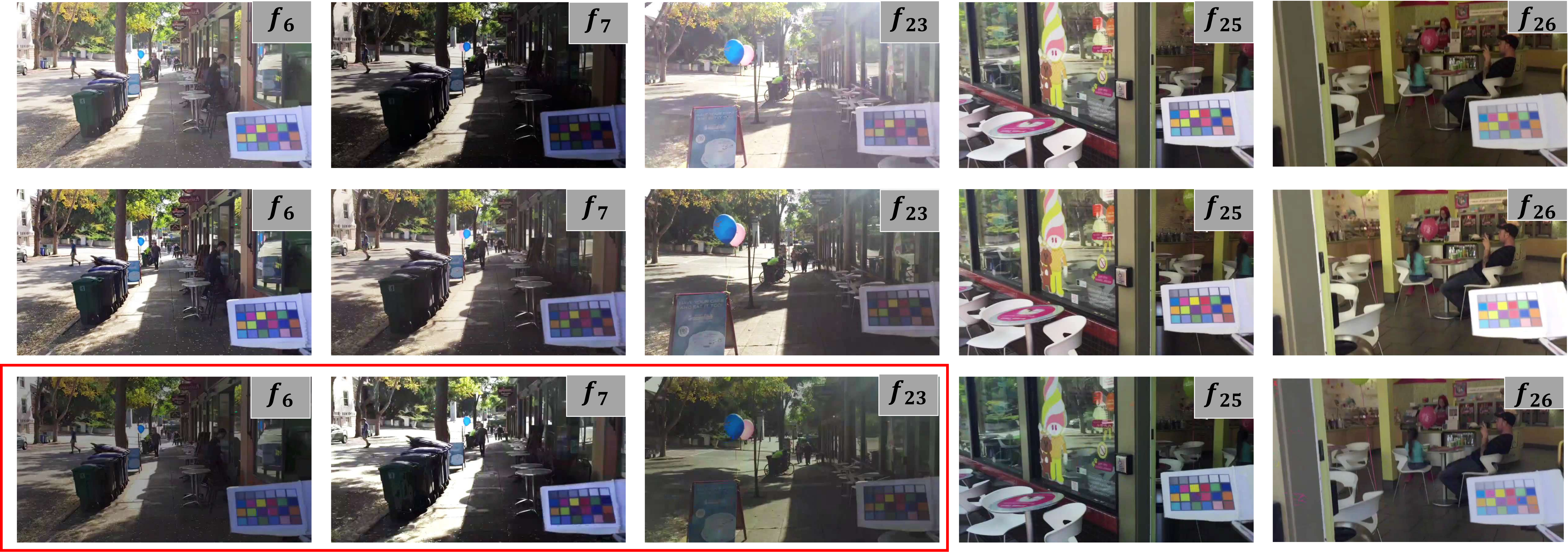}
\end{center}
   \caption{Experiment on a real video dataset. (TOP) input frame sequence; (MIDDLE) corrected frames using our method; (BOTTOM) corrected frames using Farbman \etal~\cite{farbman2011tonal}. The input frame sequence contain brightness fluctuation due to the change from outdoor to indoor. Farbman \etal~\cite{farbman2011tonal} fails to handle the over-exposed frame like $f_{23}$.}
\label{fig:result_euclid}
\end{figure*}

\para{Real videos} We also collected video datasets and applied our frame sampling (described in \Sref{sec:frame_sampling}) to generate $16\times$ fast-forward videos. These videos include an outdoor hiking sequence (\Fref{fig:result_hike}) which contains both smooth scene illumination dynamics due to spatial changes in location, and high frequency tone jitter due to camera auto-adjustments (see $f_1$ in the top row), and another challenging video that transitions from outdoor to indoor (see \Fref{fig:result_euclid}).

\para{Comparisons}
We compare our method with that from Farbman \etal~\cite{farbman2011tonal} using code released by the original authors. For each of our videos, we manually choose an anchor frame that is of good image quality. While the authors claim that multiple anchor frames can be used, selecting these frames is a tedious manual task that is not practical for real videos. \Fref{fig:synthetic_adobe} and \Fref{fig:synthetic_ts} compare the two methods on synthetic video sequences and \Fref{fig:result_hike} and \Fref{fig:result_euclid} on two real videos. Across all these comparisons, our results remove high-frequency jitter while naturally adapting to changes in content and scene illumination. In contrast, Farbman \etal. try to match the appearance in all the frames leading to unnatural looking results and significant image artifacts (especially due to errors in the dense pixel correspondence that they rely on).

We tried to compare with Wang \etal~\cite{wang2014video} using the authors' code, but their method failed during the registration step since they assume neighboring frames to have little content variation, and their method is based on the quality of registration, while few correspondences are available across frames in our test videos. We also experimented with a global affine smoothing method, which is a factorization-based algorithm we implemented. Instead of factorizing the transformation matrices using $PCA$ as in the work of Wang~\etal~\cite{wang2014video}, we decompose each affine transformation into 4 components: rotation, translation, shear and scale to avoid asynchronously smoothing different matrix entries. We then apply $L1$ smoothing to calculate a smooth path for each of the 4 components, warp the original value to its smooth path and reconstruct an affine transformation from the 4 warped components. This can be thought of as a version of \cite{wang2014video} with our robust motion estimation and pair-wise transformation computation. However, we found the parameters of global smoothing hard to control, and the smoothed transformations introduce color artifacts (see \Fref{fig:synthetic_adobe}, bottom).

\Fref{fig:synthetic_ts} shows the result on synthetic outdoor scene. In this dataset, distant background scene does not change a lot, but moving crowd makes local correspondence matching unreliable. Farbman~\etal~\cite{farbman2011tonal} suffers from local artifact and incorrect adjustment due to inaccurate local correspondences, while our photometric stabilization effectively attenuates fluctuations in color and brightness.

\begin{figure}[h!]
\begin{center}
		\includegraphics[width=1.0\linewidth]{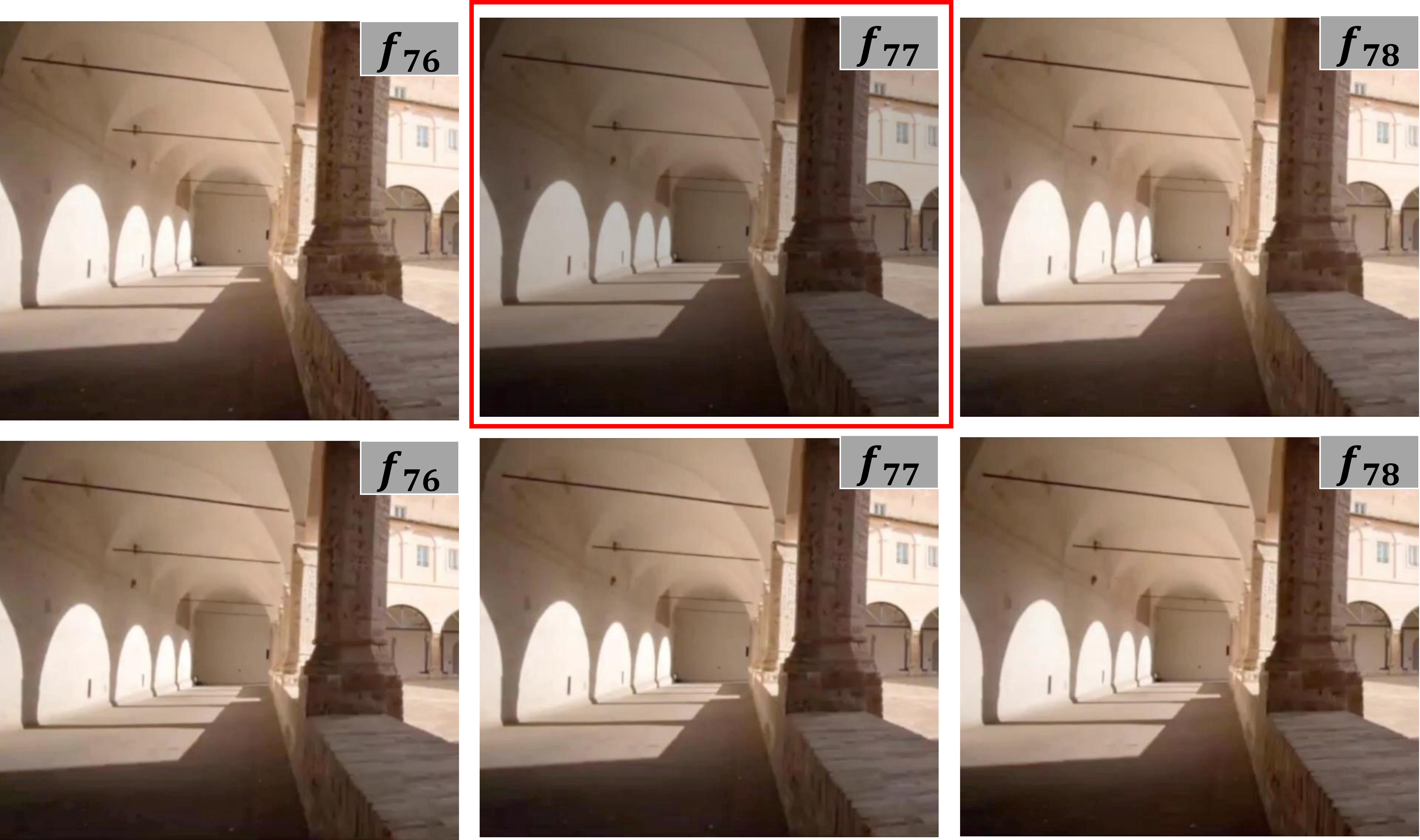}
\end{center}
	\caption{Test on timelapse videos. (TOP) original frame sequence (BOTTOM) corrected frames using our proposed method.}
\label{fig:timelapse}
\end{figure}

\para{Extension to time-lapse videos}
Our technique can also be applied to stabilize timelapse videos with large time span. Timelapse videos are usually captured from a static camera, but it may also have high frequency photometric caused by sudden illumination changes. In the example shown in \Fref{fig:timelapse}, the brightness fluctuation comes from clouds in the sky getting in and out of the frame. Our photometric stabilization can distinguish high frequency jitter from gradual illumination changes and produce a result the jitter-free time-lapse videos.

\para{Running time}
Given a 100-frame speed-up video (i.e., 8$\times$ speedup from a $800$-frame video) of resolution $1080\times1920$ (captured by an iPhone 6), our MATLAB single-thread implementation takes $212s$ to stabilize the entire sequence with 60\% of the time spent on correspondence matching. However, because our technique relies on local temporal processing, the pair-wise correspondences and transformations can computed in parallel leading to significant time gain. In comparison, the method from Farbman \etal~\cite{farbman2011tonal} takes $1957s$ to process the same test sequence with one anchor frame.
%---------------------------------------------------------------

\subsection{Conclusions}
In this paper, we have presented a photometric stabilization method for fast-forward videos. Given a video input with a desired speedup factor, we perform photometrically optimal frame selection, and then apply stabilization on the selected frames to remove high frequency color and brightness fluctuations. Our technique is able to automatically detect and correct outlier frames and can also handle large content variations across frames without any prior information or manual anchor frame selection. The algorithm is designed to be computationally efficient and has the potential to be implemented extremely fast for real-time applications.


We evaluate our stabilization algorithm on both synthetic and real fast-forward videos with high frequency brightness and color fluctuations. We sample the original video frames using our optimal frame selection technique~\Sref{sec:frame_sampling}. We request the reviewers to see the supplementary video to evaluate the quality of our results. The supplementary material also contains more results and comparisons.

We focus on removing high-frequency color/tone variations. As a result low-frequency camera variations will not be corrected by this technique. We would like to address this in the future. We also would like to explore extensions of this framework to address the problem of photometrically aligning multiple video sequences of the same scene.

\newpage
\bibliographystyle{eg-alpha-doi}
\bibliography{egbibsample}
%-------------------------------------------------------------------------
\newpage

\end{document}